# Exact Multivariate Tests – A New Effective Principle of Controlled Model Choice

Jürgen Läuter[1,2], Maciej Rosołowski[1], and Ekkehard Glimm[3]

[1]Institute of Medical Informatics, Statistics and Epidemiology (IMISE), Medical Faculty, University of Leipzig, Härtelstr. 16-18, 04107 Leipzig, Germany

[2]Otto von Guericke University Magdeburg, Mittelstr. 2/151, 39114 Magdeburg, Germany

[3]Statistical Methodology, Novartis Pharma AG, Basel, Switzerland

Address for correspondence:
Jürgen Läuter, Mittelstr. 2/151, 39114 Magdeburg, Germany
E-mail: juergen.laeuter@med.ovgu.de

Funding: This work was supported by the German BMBF network HaematoSys (BMBF Nr. 0315452A).

**Abstract**

High-dimensional tests are applied to find relevant sets of variables and relevant models. If variables are selected by analyzing the sums of products matrices and a corresponding mean-value test is performed, there is the danger that the nominal error of first kind is exceeded. In the paper, well-known multivariate tests receive a new mathematical interpretation such that the error of first kind of the combined testing and selecting procedure can more easily be kept. The null hypotheses on mean values are replaced by hypotheses on distributional sphericity of the individual score responses. Thus, model choice is possible without too strong restrictions. The method is presented for all linear multivariate designs. It is illustrated by an example from bioinformatics: The selection of gene sets for the comparison of groups of patients suffering from B-cell lymphomas.

# 1  Introduction

In many applications, the aim of multivariate analysis is to recognize the structure and the system of given individuals. At first, essential variables are determined or, more generally, sets of mutually similar variables that have high relevance. Another way of solution is to form linear combinations of variables, so-called scores, which are suitable for ordering and classifying the individuals.

It is well-known that the selection of variables and the determination of scores are subject to random fluctuations. These are particularly large if the number $p$ of variables is high and the sample size $n$ is low. Therefore, we apply so-called stable procedures where highly correlated variables are grouped into sets. We assume that such equalizing and stabilizing strategies are useful in medical and biological research, because compounds of similar variables are more easily interpretable and often more informative regarding disease status, characterization of populations etc. than single variables. Meinshausen (2008) has also discussed this situation.

Three testing principles of the model-based multivariate statistics are contrasted in this paper:
1. The classical tests for normally distributed data by Hotelling (1931), Wilks (1932) and Roy (1953) that are based on the least-squares estimators of the mean-value parameters.
2. The spherical tests for hypotheses on mean values that are derived from the theory of spherically and elliptically contoured distributions (Fang and Zhang, 1990): The tests by Läuter (1996) and Läuter, Glimm and Kropf (1996, 1998). These tests use linear scores with coefficients which are defined as functions of the so-called total sums of products matrices. These tests can be applied for arbitrarily high dimension $p$, even if the sample size $n$ is low.
3. A novel approach for testing the sphericity of score distributions: In this case, the score coefficients are also determined from the total sums of products matrices, but they may depend on all elements of these matrices regardless of whether they correspond to "null or non-null variables". Therefore, these tests can be applied much broader and more effectively than the mean-value tests of 2. However, they have the disadvantage that (in a strict sense) no results on the statistical parameters of the data are obtained, but only an information on the sphericity or non-sphericity of the score distributions.

The new tests of 3. are useful particularly in applications from bioinformatics. In gene expression analysis, very high values of the dimension *p* occur, for example $p \approx 20\,000$, but the sample sizes *n* remain mostly low, for example $n \approx 100$. The new method facilitates the derivation of linear scores that are well adapted to the given scientific problem. The method overcomes limitations of the spherical mean-value tests of 2. regarding the selection of variables. This is illustrated in Section 4. The new tests are mathematically exact in every case, a spherical score distribution is rejected with probability $\alpha$, only. If, for example, the multivariate two-groups situation is considered, then the rejection of the score sphericity means that the *n* individual score values are able to discriminate between the groups. However, in general, conclusions on mean-value differences of the selected variables cannot be drawn with this approach. Model choice and mean-value statements collide.

The score-distribution tests of 3. can also be carried out as resampling tests. In this case, the null hypothesis of permutational invariance of the score distribution is investigated instead of the null hypothesis of sphericity.

If several linear scores are considered, the tests are performed in such a way that the family-wise type I error is kept strictly.

Since many years, we know the exact multivariate tests for distribution parameters that are based on permutation and rotation principles. Westfall and Young (1993) developed selection strategies which use permutational methods and multiple testing procedures. Langsrud (2005) applied data rotations instead of permutations. However, these tests are not directly concerned with the aim of this paper.

There are different multivariate tests based on methods of asymptotic statistics, for example, the tests by Dempster (1958, 1960), Bai and Saranadasa (1996), Goeman, van de Geer, de Kort and van Houwelingen (2004), and by Srivastava and Du (2007). These approaches are not considered in this paper.

## 2  Basic methods: Classical and spherical mean-value tests

### 2.1  Classical multivariate tests

The well-known classical multivariate tests refer to an $n \times p$ data matrix **X** that consists of *n* independent and normally distributed rows $\mathbf{x}'_{(j)}$:

$$\mathbf{x}'_{(j)} \sim \mathrm{N}_p(\boldsymbol{\mu}'_{(j)}, \boldsymbol{\Sigma}), \quad j = 1, \ldots, n.$$

The *n* rows have the same covariance matrix $\boldsymbol{\Sigma}$. We assume that $\boldsymbol{\Sigma}$ is positive definite.

**One-group design**

In the special case of the one-group experimental design, the *n* mean vectors $\boldsymbol{\mu}'_{(j)}$ are equal:

$$\boldsymbol{\mu}'_{(j)} = \boldsymbol{\mu}', \quad j = 1, \ldots, n.$$

Then $\boldsymbol{\mu}'$ is estimated by $\bar{\mathbf{x}}' = \frac{1}{n} \sum_{j=1}^{n} \mathbf{x}'_{(j)} = \frac{1}{n} \mathbf{1}'_n \mathbf{X}$, and the residuals are determined on the basis of $\mathbf{X} - \bar{\mathbf{X}}$, where $\bar{\mathbf{X}} = \mathbf{1}_n \frac{1}{n} \mathbf{1}'_n \mathbf{X}$ is the mean value matrix. Here $\mathbf{1}_n$ denotes the vector consisting of *n* elements 1.

The one-group null hypothesis
$$H_0: \boldsymbol{\mu}' = \boldsymbol{0}$$
can be tested for $n-1 \geq p$ by the beta test
$$B = n\,\bar{\mathbf{x}}'(\mathbf{X}'\mathbf{X})^{-1}\bar{\mathbf{x}} \geq B_{1-\alpha}(\frac{p}{2}, \frac{n-p}{2})$$
or the equivalent $F$ test
$$F = \frac{n-p}{p}\, n\bar{\mathbf{x}}'\mathbf{G}^{-1}\bar{\mathbf{x}} \geq F_{1-\alpha}(p, n-p),$$
where $\mathbf{G} = (\mathbf{X}-\bar{\mathbf{X}})'(\mathbf{X}-\bar{\mathbf{X}})$ is the $p \times p$ residual sums of products matrix.

**General design**
In the general linear model, the classical multivariate test is determined by two $n \times n$ projection matrices $\mathbf{Q}$ of rank $f$ and $\mathbf{Q}_H$ of rank $f_H$, i.e. with $\mathbf{Q}' = \mathbf{Q} = \mathbf{Q}^2$, $\mathbf{Q}'_H = \mathbf{Q}_H = \mathbf{Q}_H^2$. The matrix $\mathbf{Q} - \mathbf{Q}_H$ must be positive semidefinite, and the ranks must fulfill the conditions $1 \leq f_H < f \leq n$, $p \leq f - f_H$. Matrix $\mathbf{Q}$ defines the subspace of the full $n$-dimensional space, in which the statistical decision takes place, and matrix $\mathbf{Q}_H$, the so-called hypothesis matrix, states the acceptance and the rejection area of the null hypothesis within the subspace.

Under the null hypothesis in the general linear model,
$$H_0: \mathbf{Q}\,E(\mathbf{X}) = \mathbf{0}$$
is fulfilled, where $E(\mathbf{X})$ is the expectation of $\mathbf{X}$. The test corresponding to $\mathbf{Q}_H$ with using the Wilks determinant criterion is
$$Lambda = \frac{|\mathbf{X}'(\mathbf{Q}-\mathbf{Q}_H)\mathbf{X}|}{|\mathbf{X}'\mathbf{Q}\mathbf{X}|} \leq \Lambda_\alpha(p, f_H, f-f_H).$$
The details on the distribution of this statistic can be found in Anderson (2003, Section 8.4).

In the special case of the one-group design, the definitions $\mathbf{Q} = \mathbf{I}_n$, $\mathbf{Q}_H = \mathbf{1}_n \frac{1}{n}\mathbf{1}'_n$, $f = n$, $f_H = 1$ are valid, where $\mathbf{I}_n$ denotes the $n \times n$ identity matrix.

**2.2 Spherical tests for mean-value hypotheses**
The so-called spherical tests utilize the property of sphericity of the normal distribution, i.e. the property that a matrix $\mathbf{X}$ consisting of $n$ independent rows $\mathbf{x}'_{(j)} \sim N_p(\mathbf{0}, \boldsymbol{\Sigma})$, $j = 1, \ldots, n$, maintains its distribution if it is transformed by any $n \times n$ orthogonal matrix $\mathbf{C}'$:
$$\mathbf{C}'\mathbf{X} \stackrel{d}{=} \mathbf{X}.$$
In the case of the spherical distribution, the variables 1 to $p$ may be linearly combined into scores which are again spherically distributed.

**Spherical mean-value tests in the one-group design**
If the null hypothesis $\boldsymbol{\mu}' = \mathbf{0}$ is fulfilled in this design, $\mathbf{X}$ is spherically distributed. Then a $p \times q$ weight matrix $\mathbf{D}$ defined as a function of the $p \times p$ sums of products matrix $\mathbf{X}'\mathbf{X}$ generates an $n \times q$ score matrix $\mathbf{Z} = \mathbf{X}\mathbf{D}$, which is also spherically distributed. In the spherical test, $\mathbf{Z}$ is analyzed instead of $\mathbf{X}$. In case of $q < p$, the dimension of the testing problem is reduced via the transformation from $\mathbf{X}$ to $\mathbf{Z}$. Thus, extremely large dimensions $p$ can be compressed into very small dimensions $q$.

**X** is supposed to consist of *n* independent rows $\mathbf{x}'_{(j)} \sim \mathrm{N}_p(\boldsymbol{\mu}', \boldsymbol{\Sigma})$, $j = 1, \ldots, n$. The *p*-dimensional null hypothesis $\boldsymbol{\mu}' = \mathbf{0}$ is assessed by the beta test

$$\mathrm{B} = n\,\bar{\mathbf{z}}'(\mathbf{Z}'\mathbf{Z})^{-1}\bar{\mathbf{z}} = n\,\bar{\mathbf{x}}'\mathbf{D}(\mathbf{D}'\mathbf{X}'\mathbf{X}\mathbf{D})^{-1}\mathbf{D}'\bar{\mathbf{x}} \;\geq\; \mathrm{B}_{1-\alpha}(\frac{q}{2}, \frac{n-q}{2})$$

(Läuter, 1996; Läuter et al., 1996, 1998). Here, we assume that $n - 1 \geq q$ and **Z** has rank *q* with probability 1. A significance means that the sphericity of the **Z** distribution and, consequently, the null centrality of the **X** distribution is rejected. This test keeps the level $\alpha$ exactly.

However, using this testing principle as a basis for selecting variables requires the weight matrix **D** to be defined only by those **X** variables which are being analyzed. In the case of

$\mathbf{X} = (\mathbf{X}_1 \quad \mathbf{X}_2)$, $\boldsymbol{\mu}' = (\boldsymbol{\mu}'_1 \quad \boldsymbol{\mu}'_2)$, $\mathbf{D} = \begin{pmatrix} \mathbf{D}_1 \\ \mathbf{0} \end{pmatrix}$, if the hypothesis $\boldsymbol{\mu}'_1 = \mathbf{0}$ is tested with the score $\mathbf{Z}_1 = \mathbf{X}_1 \mathbf{D}_1$, then $\mathbf{D}_1$ must be uniquely determined by the submatrix $\mathbf{X}'_1 \mathbf{X}_1$ of $\mathbf{X}'\mathbf{X}$. To overcome this restriction, a new proposal of testing is presented subsequently in Section 3.

**Test procedure by Kropf for searching non-null variables in the one-group design**
In the context of the spherical mean-value tests, Kropf has developed a multiple testing procedure for searching single variables with mean values $\mu_i \neq 0$ (Kropf, 2000; Kropf and Läuter, 2002). This method uses weight vectors $\mathbf{d}_1, \ldots, \mathbf{d}_p$ which contain one element 1 and $p - 1$ elements 0. The vector $\mathbf{d}_1$ has 1 in the position $i_1$ corresponding to the largest diagonal element of $\mathbf{X}'\mathbf{X}$. In $\mathbf{d}_2$, element 1 is in position $i_2$, corresponding to the second-largest diagonal element of $\mathbf{X}'\mathbf{X}$, and so on. Thus, an ordered sequence of univariate beta tests is obtained:

$$\mathrm{B}_h = \frac{n\,(\bar{\mathbf{x}}'\mathbf{d}_h)^2}{\mathbf{d}'_h \mathbf{X}'\mathbf{X}\mathbf{d}_h} = \frac{n\,\bar{x}^2_{i_h}}{\sum_{j=1}^n x^2_{ji_h}} \;\geq\; \mathrm{B}_{1-\alpha}(\frac{1}{2}, \frac{n-1}{2}), \quad h = 1, \ldots, p.$$

If we assume that at least one variable with $\mu_i = 0$ is present, then the sorting $i = i_1, \ldots, i_p$ given by matrix $\mathbf{X}'\mathbf{X}$ determines uniquely a null variable as the first one.

In Kropf's procedure, the ordered beta tests $\mathrm{B}_1, \mathrm{B}_2, \mathrm{B}_3, \ldots$ are performed as long as significances result. The procedure stops at the first non-significant test. The significance of $\mathrm{B}_h$ is interpreted as the proof of $\mu_{i_h} \neq 0$. No further $\alpha$-adjustment is necessary. The multiple procedure keeps the familywise type I error $\alpha$ in the strong sense (i.e. without any further assumptions on the mean values).

For this procedure, it is essential that the sorting of the null variables is not influenced by the non-null variables. This is attained by using only the diagonal elements of $\mathbf{X}'\mathbf{X}$ for ordering the tests. Hence, the correlations of the variables do not play an explicit role. Of course, the procedure is not scale-invariant.

In Section 3.2, Example 2, a similar procedure is introduced which, however, does not provide results on mean values.

**Spherical mean-value tests for principal components in the one-group design**

To illustrate the limits of spherical tests, we will also apply this method to arbitrary non-null variables with the aim to find linear combinations, which have a mean value different from zero. We begin with $n$ independent rows $\mathbf{x}'_{(j)} \sim N_p(\boldsymbol{\mu}', \boldsymbol{\Sigma})$, $j = 1, \ldots, n$, of matrix $\mathbf{X}$, where $\boldsymbol{\mu}'$ can have an arbitrary value. The solution of the eigenvalue problem

$$(\mathbf{X'X})\mathbf{d} = \mathbf{d}\lambda, \quad \mathbf{d} \neq \mathbf{0}, \quad \lambda > 0,$$

provides the coefficient vector of a principal component score $\mathbf{z} = \mathbf{Xd}$. We prove that significance of the beta test

$$B = \frac{n(\bar{\mathbf{x}}'\mathbf{d})^2}{\mathbf{d'X'Xd}} \geq B_{1-\alpha}(\frac{1}{2}, \frac{n-1}{2})$$

enables the conclusion $\boldsymbol{\mu}'\mathbf{d} \neq 0$. This means that the principal component of $\mathbf{X}$ variables defined by $\mathbf{d}$ has a mean value unequal to zero. We stress that this method cannot be extended to arbitrary other random weights $\mathbf{d}$ that are defined by $\mathbf{X'X}$.

It is obvious from the spherical tests that the significance of $B$ leads to rejection of the null hypothesis $\boldsymbol{\mu}' = \mathbf{0}$. However, $\boldsymbol{\mu}'\mathbf{d} \neq 0$ does not follow from this.

To show that we can nevertheless conclude this in the special case of the principal component $\mathbf{z} = \mathbf{Xd}$, let us assume now $n \geq 2$, $p \geq 2$, $\boldsymbol{\mu}' \neq \mathbf{0}$. For the proof, we transform the data matrix $\mathbf{X}$ into $\mathbf{Y} = \mathbf{X}(\mathbf{I}_p - \boldsymbol{\mu}(\boldsymbol{\mu}'\boldsymbol{\mu})^{-1}\boldsymbol{\mu}')$. This matrix consists of $n$ independent rows

$$\mathbf{y}'_{(j)} \sim N_p(\mathbf{0}, (\mathbf{I}_p - \boldsymbol{\mu}(\boldsymbol{\mu}'\boldsymbol{\mu})^{-1}\boldsymbol{\mu}')\boldsymbol{\Sigma}(\mathbf{I}_p - \boldsymbol{\mu}(\boldsymbol{\mu}'\boldsymbol{\mu})^{-1}\boldsymbol{\mu}')), \quad j = 1, \ldots, n,$$

with the expectation $\mathbf{0}$. In practice, we do not know $\mathbf{Y}$, because $\boldsymbol{\mu}'$ is unknown. However, this is unimportant for the proof. Matrix $\mathbf{Y}$ is spherically distributed.

Consider the distributions of $\mathbf{Y}$ and $\mathbf{X}$ conditional on $\mathbf{Y'Y} = \text{const}$. Additionally, a fixed value $\lambda^* > 0$ is assumed. In the following, we will show that, with the probability $\alpha$, at most, we can find any weight vector $\mathbf{d} \neq \mathbf{0}$ fulfilling simultaneously the equations $(\mathbf{X'X})\mathbf{d} = \mathbf{d}\lambda^*$ and $\boldsymbol{\mu}'\mathbf{d} = 0$ with significance in the above $B$ test: If we suppose that a vector $\mathbf{d} \neq \mathbf{0}$ is given with $(\mathbf{X'X})\mathbf{d} = \mathbf{d}\lambda^*$ and $\boldsymbol{\mu}'\mathbf{d} = 0$, then it follows

$$\mathbf{Yd} = \mathbf{Xd}$$

and

$$\mathbf{Y'Yd} = \mathbf{Y'Xd} = (\mathbf{I}_p - \boldsymbol{\mu}(\boldsymbol{\mu}'\boldsymbol{\mu})^{-1}\boldsymbol{\mu}')\mathbf{X'Xd} = (\mathbf{I}_p - \boldsymbol{\mu}(\boldsymbol{\mu}'\boldsymbol{\mu})^{-1}\boldsymbol{\mu}')\mathbf{d}\lambda^* = \mathbf{d}\lambda^*.$$

Since $\mathbf{Y'Y}$ has multiple positive eigenvalues only with probability 0, $\mathbf{d}$ is the uniquely determined eigenvector of $\mathbf{Y'Y}$ belonging to the eigenvalue $\lambda^*$ (apart from differences in the normalization of the eigenvector). This rule is valid for an arbitrary mean vector $\boldsymbol{\mu}'$.

The conditional spherical beta test

$$B = \frac{n(\bar{\mathbf{y}}'\mathbf{d})^2}{\mathbf{d'Y'Yd}} \geq B_{1-\alpha}(\frac{1}{2}, \frac{n-1}{2})$$

applied to the fixed vector $\mathbf{d}$ yields significance exactly with probability $\alpha$. For each $\mathbf{X}$ with $(\mathbf{X'X})\mathbf{d} = \mathbf{d}\lambda^*$ in the conditional distribution $\mathbf{Y'Y} = \text{const}.$, we obtain $\mathbf{Xd} = \mathbf{Yd}$ from $\boldsymbol{\mu}'\mathbf{d} = 0$, and the corresponding beta result

$$B = \frac{n(\bar{\mathbf{x}}'\mathbf{d})^2}{\mathbf{d'X'Xd}}$$

coincides with the beta of the **Y** test. However, the level of significance α is not always exhausted in this case, because $(\mathbf{X}'\mathbf{X})\mathbf{d} = \mathbf{d}\lambda^*$ is not valid for all **X** in the conditional distribution. These considerations are sufficient for the proof.

In the practical application of the test, we will use any eigenvector $\mathbf{d}_h$ with its eigenvalue $\lambda_h$ corresponding to the first, second, third, … principal component. The given proof secures that the level of significance is kept conditionally for the fixed values of $\mathbf{Y}'\mathbf{Y}$ and $\lambda_h$.

This method of constructing the test cannot easily be applied to arbitrary other random linear combinations $\mathbf{z} = \mathbf{X}\mathbf{d}$. For the proof, it is important that the weight vector **d** defined as a function of $\mathbf{X}'\mathbf{X}$, which fulfills the relation $\boldsymbol{\mu}'\mathbf{d} = 0$, can also be represented by the same function from $\mathbf{Y}'\mathbf{Y}$. Here, $\boldsymbol{\mu}'$ is considered as arbitrary and unknown.

**Spherical mean-value tests in the general design**
We assume a matrix **X** that consists of *n* independent rows $\mathbf{x}'_{(j)} \sim N_p(\boldsymbol{\mu}'_{(j)}, \boldsymbol{\Sigma})$ with the same covariance matrix. The spherical test refers to the mean-value parameters $\boldsymbol{\mu}'_{(j)}$. If the experimental design is given by the projection matrices **Q** of rank *f* and $\mathbf{Q}_H$ of rank $f_H$, where $\mathbf{Q} - \mathbf{Q}_H$ is positive semidefinite and $1 \le f_H < f \le n$ is fulfilled, then the $p \times q$ weight matrix **D** is defined as a function of sums of products matrix $\mathbf{X}'\mathbf{Q}\mathbf{X}$. The $n \times q$ score matrix $\mathbf{Z} = \mathbf{Q}\mathbf{X}\mathbf{D}$ is introduced, where $q \le f - f_H$ and **Z** has rank *q* with probability 1. Under the null hypothesis, the condition
$$H_0 : \mathbf{Q}\,E(\mathbf{X}) = \mathbf{0}$$
is valid. The spherical test corresponding to the hypothesis matrix $\mathbf{Q}_H$ is
$$Lambda = \frac{|\mathbf{Z}'(\mathbf{Q} - \mathbf{Q}_H)\mathbf{Z}|}{|\mathbf{Z}'\mathbf{Z}|} = \frac{|\mathbf{D}'\mathbf{X}'(\mathbf{Q} - \mathbf{Q}_H)\mathbf{X}\mathbf{D}|}{|\mathbf{D}'\mathbf{X}'\mathbf{Q}\mathbf{X}\mathbf{D}|} \le \Lambda_\alpha(q, f_H, f - f_H).$$
In the case of $q = 1$, the score vector $\mathbf{z} = \mathbf{Q}\mathbf{X}\mathbf{d}$ arises, and the corresponding spherical test is
$$B = \frac{\mathbf{z}'\mathbf{Q}_H\mathbf{z}}{\mathbf{z}'\mathbf{z}} = \frac{\mathbf{d}'\mathbf{X}'\mathbf{Q}_H\mathbf{X}\mathbf{d}}{\mathbf{d}'\mathbf{X}'\mathbf{Q}\mathbf{X}\mathbf{d}} \ge B_{1-\alpha}(\frac{f_H}{2}, \frac{f - f_H}{2}).$$

## 3 The new proposal of testing score distributions for sphericity

### 3.1 General characterization
Section 2.2 has shown that the spherical mean-value tests may be used for the selection of variables only under strong conditions concerning the choice of weights **D**. Especially in applications with a high dimension *p*, where radical and complicated data compressions are necessary, these conditions are too restrictive for an effective, exact multivariate analysis. Therefore, we suggest new test strategies that do not try to investigate the mean values of the **X** variables but rather consider the distributions of linear scores **Z** constructed from the observed data. In our statistical analysis, the thinking in notions of the original variables **X** is replaced by an operational thinking on the data-structural effects of the scores **Z**. However, the scores **Z** have only a stochastic link with the original variables **X**. Hence, statements about **Z** are not in a uniquely fixed relationship with the statements on **X** variables.

## 3.2 Score-distribution tests in the one-group design

Let us consider fixed values of *n* and *p* and a family of distributions of $n \times p$ matrices **X**. The **X** distributions belonging to the family can have arbitrary properties. Each $(np)$-dimensional distribution is admitted. We demand neither normal distribution nor independence of the rows of **X**. We do also not need the characterization of the variables as null or non-null variables. The tests which are being formulated have the aim to separate "good" and "bad" distributions in the family of **X** distributions.

In the one-group design, the $p \times q$ weight matrix
$$\mathbf{D} = (\mathbf{d}_1 \quad \mathbf{d}_2 \quad \ldots \quad \mathbf{d}_q)$$
is defined as a function of the sums of products matrix $\mathbf{X}'\mathbf{X}$. The *n*-dimensional score vectors
$$\mathbf{z}_h = \mathbf{X}\mathbf{d}_h, \qquad h = 1, \ldots, q,$$
are checked for sphericity of their distributions by the beta tests
$$\mathrm{B}_h = \frac{n\, \bar{z}_h^2}{\mathbf{z}_h' \mathbf{z}_h} \;\geq\; \mathrm{B}_{1-\alpha}(\tfrac{1}{2}, \tfrac{n-1}{2}), \qquad h = 1, \ldots, q.$$
Here, the mean values
$$\bar{z}_h = \frac{1}{n} \sum_{j=1}^{n} z_{jh} = \frac{1}{n} \mathbf{1}_n' \mathbf{z}_h, \qquad h = 1, \ldots, q,$$
are applied. Each distribution of the **X** family is supposed to fulfill the regularity condition that both the definition of **D** and the beta tests for sphericity of the scores $\mathbf{z}_1, \ldots, \mathbf{z}_q$ are possible and unique with probability 1.

We use the following null hypotheses for the scores $\mathbf{z}_h$, $h = 1, \ldots, q$:

$H_{0h}$: The distribution of **X** generates a score $\mathbf{z}_h$ which is spherically distributed,

i.e. $\mathbf{C}' \mathbf{z}_h \stackrel{d}{=} \mathbf{z}_h$ for each fixed orthogonal matrix $\mathbf{C}'$.

The null hypothesis means that the distribution of $\mathbf{z}_h$ is invariant against arbitrary rotations and reflections in the *n*-dimensional space. This implies that all directions of the *n*-dimensional space are equally likely.

As mentioned in the introduction, the hypotheses $H_{0h}$ are not formulated as hypotheses about parameters of the distributions of $\mathbf{z}_h$ or **X**. This is in the spirit of Lehmann (1975), Chapter 1, who also emphasizes that in non-parametric statistics, the hypothesis of "no treatment effect" is usually a statement about a property of distributions which cannot be converted into a statement about corresponding parameters. In the characterization of Liang and Fang (2000, p. 926), the high-dimensional behaviour of **X** is analyzed over one-dimensional "projection tests" for the scores $\mathbf{z}_h$.

Under the null hypothesis $H_{0h}$, the beta test $\mathrm{B}_h$ yields significance exactly with probability $\alpha$. This is a consequence from the theory of spherical tests (Läuter, Glimm and Kropf, 1998, Theorem 2). Significance of $\mathrm{B}_h$ is obtained if the mean value $\bar{z}_h$ has a sufficiently large deviation from zero. Thus, the rejection of the sphericity null hypothesis $H_{0h}$ also expresses that the score values $z_{jh}$, $j = 1, \ldots, n$, deviate from zero in a systematic way, corresponding to the one-group experimental design.

The beta test $B_h$ can also be considered as a conditional test for spherical distribution of the score $z_h$ under the condition $X'X = \text{const}$. Then $d_h$ and $z'_h z_h = d'_h X'X d_h$ have fixed values.

For the new testing method, it is important that $X$ distributions exist for each defined function of the weights $d$ (possibly outside the given family of $X$ distributions), which lead to spherically distributed scores $z = Xd$ and, correspondingly, to the exact rejection probability $\alpha$. For example, spherically distributed scores $z$ arise for an arbitrary $n \times p$ left-spherical matrix distribution of $X$ (Fang and Zhang, 1990). These scores $z$ attain significance exactly with probability $\alpha$. Special distributions of this kind are the $X$ distributions formed by $n$ independent rows $x'_{(j)} \sim N_p(0, \Sigma)$. Depending on the given definition of $d$, further $X$ distributions can also exist that lead to a spherical distribution of $z$.

On this basis, a multiple testing procedure can be applied to the sequence of scores $z_1, z_2, z_3, \ldots$, which works similarly as Kropf's procedure of Section 2.2. The ordered beta tests $B_1, B_2, B_3, \ldots$ are performed as long as significances on the level $\alpha$ result. The order of the scores is determined uniquely by the matrix $D$. A significance of $B_h$ is interpreted as the proof of the non-sphericity of score $z_h$. With the first non-significant test, the procedure is finished. The procedure strictly keeps the familywise type I error $\alpha$, i.e., in the series of all significant scores $z_h$, some falsely significant scores – scores with a spherical distribution instead of a non-spherical one – may occur with probability $\alpha$, at most.

Additionally, we note that the modification of the multiple testing procedure by Hommel and Kropf (2005) can also be applied. In this modification, the sequence of tests $B_1, B_2, B_3, \ldots$ is stopped only when the $k$th non-significant test is obtained. Correspondingly, the adjusted significance level $\alpha/k$ must then be used. Here, $k$ is a fixed prescribed number. We will utilize this method in Section 4.

**Some examples:**
1. The most straightforward application of the testing method is given by the principal component scores. These scores are suitable to combine the information of all $p$ variables, and they allow tests according to the one-group design. The weight vectors $d_1, d_2, \ldots, d_q$ are defined as the eigenvectors of the eigenvalue problem
$$X'X d_h = d_h \lambda_h, \quad d'_h d_h = 1,$$
pertaining to the largest eigenvalue $\lambda_1$, the second-largest eigenvalue $\lambda_2$, and so on. To attain a unique orientation of the vectors $d_h$, one can additionally demand that their first elements $d_{1h}$ must be positive. Thus, the principal component scores $z_h = X d_h$, $h = 1, \ldots, q$, are found. These scores are successively tested until the first non-significance (or the $k$th non-significance) occurs. Significance of a score $z_h$ is considered as the proof that $z_h$ has a non-spherical distribution. The significance also shows that the responses $z_{jh}$, $j = 1, \ldots, n$, of the $n$ individuals can be separated from zero.

   In this procedure, we do not need special assumptions about the distribution of $X$. In contrast to the principal component test of Section 2.2, the assumption of a normal distribution is not necessary.

   This very simple and mathematically exact testing strategy is a new, useful application of the principal components. Non-centralities of the $X$ variables are included in the tests.

Many other authors use only heuristic methods, e.g. so-called scree plots of the eigenvalues, for assessing principal components. In the practical example of Section 4.2, the principal component scores are applied.

2. As another application, we consider a method that arises from Kropf's procedure of Section 2.2 by a changed sorting rule. Single variables and corresponding scores $\mathbf{z}_1, \mathbf{z}_2, \ldots, \mathbf{z}_p$ are determined which are suitable for the characterization of individuals in the one-group model. Here, the variables are ordered with respect to the column sums $s_i = \sum_{g=1}^{p} \text{abs}(w_{gi})$ of the sums of products matrix $\mathbf{W} = \mathbf{X}'\mathbf{X}$. If $s_{i_1}$ is the largest column sum of $\mathbf{W}$, then $\mathbf{z}_1$ is equated to column $i_1$ of $\mathbf{X}$. If $s_{i_2}$ is the second-largest column sum, then $\mathbf{z}_2$ is equated to column $i_2$ of $\mathbf{X}$, and so on. This means that the score vectors are determined from the columns $\mathbf{x}_i$ of $\mathbf{X}$, and the assignment rule is defined by matrix $\mathbf{W}$.

Significance of a score $\mathbf{z}_h$ means that $\mathbf{z}_h$ can be considered as non-spherically distributed, i.e., this score vector enables the statistically controlled characterization of the $n$ individuals. Though the computation of $\mathbf{z}_h$ is based on a single column $\mathbf{x}_{i_h}$ of $\mathbf{X}$, the test does not allow confirmatory conclusions about the mean value $\mu_{i_h}$ of column $\mathbf{x}_{i_h}$, because the index $i_h$ depends on all $\mathbf{X}$ columns in a random manner.

In certain cases, the score $\mathbf{z}_h$ can be significant with a probability larger than $\alpha$, even if it only corresponds to columns with mean values $\mu_{i_h} = 0$. To demonstrate this, we consider the $10 \times 3$ matrix $\mathbf{X}$ with independent rows $\mathbf{x}'_{(j)} \sim N_3((0 \ 0 \ 3), \mathbf{I}_3)$, $j = 1, \ldots, 10$. The computer simulation with $10^6$ runs for $\alpha = 0.05$ yields the significant first score $\mathbf{z}_1 = \mathbf{x}_3$ almost with probability 1 due to the mean value 3 in $\mathbf{x}_3$. The second score $\mathbf{z}_2$, significant with probability $0.0834$, is equal to $\mathbf{x}_1$ or $\mathbf{x}_2$. This probability is higher than $0.05$ because in stochastic tendency, $\mathbf{z}_2$ is more often identified as the $\mathbf{x}_1$ or $\mathbf{x}_2$ that randomly has the larger mean value. Therefore, a corresponding mean-value test with multiple error control in the strong sense is impossible on this basis (only the control in the weak sense is given). However, the method is correct for testing sphericity. The level $\alpha$ is exceeded only if the score distribution is non-spherical.

Contrary to Kropf's procedure of Section 2.2, we do not need confining assumptions on the distribution of $\mathbf{X}$, to attain exact statements on the non-sphericity of the scores $\mathbf{z}_h$.

3. Consider random $n \times 2$ matrices $\mathbf{X} = (\mathbf{x}_1 \ \mathbf{x}_2)$. We define the weight vector
$$\mathbf{d} = \begin{pmatrix} -\dfrac{\mathbf{x}'_1 \mathbf{x}_2}{\mathbf{x}'_1 \mathbf{x}_1} \\ 1 \end{pmatrix}$$
and the "regression score" $\mathbf{z} = \mathbf{X}\mathbf{d} = \mathbf{x}_2 - \mathbf{x}_1 \dfrac{\mathbf{x}'_1 \mathbf{x}_2}{\mathbf{x}'_1 \mathbf{x}_1}$. This score vector shows whether variable 2 can be replaced by variable 1 in the one-group design, whether variable 2 is "redundant" with respect to variable 1. We will express the property of redundancy by means of the spherical distribution of the score vector $\mathbf{z}$. Thus, the test for redundancy

$$B = \frac{n \bar{z}^2}{\mathbf{z}'\mathbf{z}} \geq B_{1-\alpha}(\frac{1}{2}, \frac{n-1}{2})$$

is obtained. Significance in this test can be interpreted as the proof of non-redundancy of variable 2.

An additional justification of this test is found if the special case of a matrix **X** consisting of $n$ independent normally distributed rows

$$\mathbf{x}'_{(j)} \sim N_2((\mu_1 \quad \mu_2), \begin{pmatrix} \sigma_{11} & \sigma_{12} \\ \sigma_{12} & \sigma_{22} \end{pmatrix})$$

is considered. In this case, the theory by Rao (1948, 1973) yields the redundancy condition $\mu_2 - \mu_1 \frac{\sigma_{12}}{\sigma_{11}} = 0$ (see also Timm (2002, p. 242) and Seber (2004, p. 471)). Then, the beta test leads to significance with a probability smaller than or equal to $\alpha$. If even $\mu_1 = \mu_2 = 0$, the assumptions of the spherical mean-value test of Section 2.2 are fulfilled, so that the rejection rate $\alpha$ is attained precisely.

In the case that a matrix $\mathbf{X}_1$ of several columns is given instead of $\mathbf{x}_1$, i.e. $\mathbf{X} = (\mathbf{X}_1 \quad \mathbf{x}_2)$, the regression score $\mathbf{z} = \mathbf{x}_2 - \mathbf{X}_1(\mathbf{X}'_1\mathbf{X}_1)^{-1}\mathbf{X}'_1\mathbf{x}_2$ can be applied to check the redundancy of $\mathbf{x}_2$ with respect to $\mathbf{X}_1$.

### 3.3 Score-distribution tests in the two-groups design and the correlation design

In the case of the two-groups experimental design, the set of $n$ individuals is divided into two groups with $n^{(1)}$ and $n^{(2)}$ individuals. The aim of our analysis is to test the separability of the groups. Then, the data matrix has the shape

$$\mathbf{X} = \begin{pmatrix} \mathbf{X}^{(1)} \\ \mathbf{X}^{(2)} \end{pmatrix}.$$

Here, the weight matrix $\mathbf{D} = (\mathbf{d}_1 \quad \mathbf{d}_2 \quad \ldots \quad \mathbf{d}_q)$ is introduced as a function of the sums of products matrix $(\mathbf{X} - \overline{\mathbf{X}})'(\mathbf{X} - \overline{\mathbf{X}})$, where the matrix of the total mean values

$$\overline{\mathbf{X}} = \mathbf{1}_n \frac{1}{n} \mathbf{1}'_n \mathbf{X} = \mathbf{1}_n \overline{\mathbf{x}}'$$

is used. The score vectors are defined by

$$\mathbf{z}_h = \begin{pmatrix} \mathbf{z}_h^{(1)} \\ \mathbf{z}_h^{(2)} \end{pmatrix} = (\begin{pmatrix} \mathbf{X}^{(1)} \\ \mathbf{X}^{(2)} \end{pmatrix} - \overline{\mathbf{X}})\mathbf{d}_h, \quad h = 1, 2, \ldots, q,$$

and they are checked by the beta tests

$$B_h = \frac{n^{(1)}n^{(2)}}{n^{(1)}+n^{(2)}} \frac{(\overline{z}_h^{(1)} - \overline{z}_h^{(2)})^2}{\mathbf{z}'_h \mathbf{z}_h} \geq B_{1-\alpha}(\frac{1}{2}, \frac{n-2}{2}), \quad h = 1, 2, \ldots, q.$$

In this formula, $\overline{z}_h^{(1)}$ and $\overline{z}_h^{(2)}$ are the mean values of the subvectors $\mathbf{z}_h^{(1)}$ and $\mathbf{z}_h^{(2)}$.

The data **X** can have an arbitrary distribution. The subdivision of individuals in the groups (1) and (2) plays a role only for testing the scores $\mathbf{z}_h$. We test the null hypothesis that the score vector $\mathbf{z}_h$ is spherically distributed in the $(n-1)$-dimensional subspace of all vectors **z**, which have the total mean value $\overline{z} = \frac{1}{n}\sum_{j=1}^{n} z_j = 0$. The beta test rejects the null hypothesis when a sufficiently large difference between the mean values $\overline{z}_h^{(1)}$ and $\overline{z}_h^{(2)}$ of both groups is observed. Thus, significance of the test means that the score $\mathbf{z}_h$ can separate the groups.

The multiple procedure for the scores $\mathbf{z}_1, \mathbf{z}_2, \mathbf{z}_3, \ldots$ of Section 3.2 can also be applied in the case of the two-groups design. The examples of Section 3.2 can be adapted to the modified experimental situation.

We would like to emphasize the generality of this approach. For example, the case of $\mathbf{X}^{(1)}$ and $\mathbf{X}^{(2)}$ having different covariance matrices $\mathbf{\Sigma}^{(1)}$ and $\mathbf{\Sigma}^{(2)}$, the case of the so-called Behrens-Fisher problem, is also included. In our strategy, however, we decide only on sphericity or non-sphericity of the scores $\mathbf{z}_h$, where the different covariances are not utilized. Bennet (1951), Anderson (1963, 2003) and Yao (1965) have treated more special tests of the null hypothesis $\mathbf{\mu}^{(1)} = \mathbf{\mu}^{(2)}$ in case of $\mathbf{\Sigma}^{(1)} \neq \mathbf{\Sigma}^{(2)}$, in which the mean values and the covariances of the $\mathbf{X}$ distribution are checked in detail.

In the correlation design, an $n \times 1$ target vector $\mathbf{y}$ with $\overline{\mathbf{y}} = \mathbf{0}$ is given. The weights $\mathbf{d}_h$ are determined as above and the scores by $\mathbf{z}_h = (\mathbf{X} - \overline{\mathbf{X}})\mathbf{d}_h$. The beta sphericity tests

$$\mathrm{B}_h = \frac{(\mathbf{z}_h'\mathbf{y})^2}{(\mathbf{z}_h'\mathbf{z}_h)(\mathbf{y}'\mathbf{y})} \geq \mathrm{B}_{1-\alpha}(\frac{1}{2}, \frac{n-2}{2}), \quad h = 1, 2, \ldots, q,$$

are applied. A significance shows that $\mathbf{z}_h$ and $\mathbf{y}$ correlate. In the special case of principal components regression (see, for example, Hastie, Tibshirani and Friedman, 2001), the eigenvectors of $(\mathbf{X} - \overline{\mathbf{X}})'(\mathbf{X} - \overline{\mathbf{X}})$ are used as weight vectors $\mathbf{d}_h$. Then, the beta tests allow the recognition of the relevant principal components.

### 3.4 Score-distribution tests in the general design

Consider the general linear model with the projection matrices $\mathbf{Q}$ of rank $f$ and $\mathbf{Q}_\mathrm{H}$ of rank $f_\mathrm{H}$, where $\mathbf{Q} - \mathbf{Q}_\mathrm{H}$ is positive semidefinite and $1 \leq f_\mathrm{H} < f \leq n$ is valid. Then, the $p \times q$ weight matrix $\mathbf{D} = (\mathbf{d}_1 \ \mathbf{d}_2 \ \ldots \ \mathbf{d}_q)$ is introduced as a function of the sums of products matrix $\mathbf{X}'\mathbf{Q}\mathbf{X}$. The score vectors are defined by

$$\mathbf{z}_h = \mathbf{Q}\mathbf{X}\mathbf{d}_h, \quad h = 1, \ldots, q,$$

and they can successively be evaluated by the beta tests

$$\mathrm{B}_h = \frac{\mathbf{z}_h'\mathbf{Q}_\mathrm{H}\mathbf{z}_h}{\mathbf{z}_h'\mathbf{z}_h} \geq \mathrm{B}_{1-\alpha}(\frac{f_\mathrm{H}}{2}, \frac{f - f_\mathrm{H}}{2}), \quad h = 1, \ldots, q.$$

The null hypothesis of this test postulates that the score vector $\mathbf{z}_h$ is spherically distributed in the $f$-dimensional subspace spanned by the columns of $\mathbf{Q}$. Under this null hypothesis, the $\mathrm{B}_h$ test yields significance exactly with probability $\alpha$.

The null hypothesis can be fulfilled for any given projection matrix $\mathbf{Q}$ and any definition of the weight vector $\mathbf{d}_h$. For example, sphericity of $\mathbf{z}_h$ is obtained for a distribution $\mathbf{X}$ that consists of $n$ independent rows $\mathbf{x}_{(j)}' \sim \mathrm{N}_p(\mathbf{\mu}_{(j)}', \mathbf{\Sigma})$ and satisfies $\mathbf{Q}\,\mathrm{E}(\mathbf{X}) = \mathbf{0}$. Thus, the new testing method receives its necessary justification: It allows a distinction of "good" distributions, with deviations from the sphericity corresponding to the experimental design, and "bad" distributions, without such deviations.

The multiple procedure for the scores $\mathbf{z}_1, \mathbf{z}_2, \mathbf{z}_3, \ldots$ can be performed as described in Section 3.2.

# 4 Applications in bioinformatics

## 4.1 Gene expression data

We will consider gene expression data from the project "Molecular Mechanisms in Malignant Lymphomas Network" of the Deutsche Krebshilfe that we have analyzed in the Institute of Medical Informatics, Statistics and Epidemiology of the Leipzig University. In particular, we utilize the previously published (Hummel et al., 2006) and extensively applied data set (e.g., Bentink et al., 2008; Maneck et al., 2011; Schramedei et al. 2011) that was derived from patients with B-cell lymphomas. Here, we only use a subset of these data consisting of 114 patients which served as a training set in Hummel et al. (2006).

The gene expression data are stored on Affymetrix HGU133A microarrays which contain 22277 gene probe sets. However, we exploit only $p = 9577$ preselected genes in our computation, to avoid the multiple use of the same genes.

We will apply our new testing principle, in which scores are checked for a spherical distribution. The aim is to construct scores for the separation of two groups of patients, the first group with myc-break ($n^{(1)} = 40$), the second group without myc-break ($n^{(2)} = 70$). The four remaining patients have no myc-break assignment. Myc is a strong proto-oncogene, and it is found to be upregulated in many types of cancers. One of the mechanisms which lead to persistent activation of myc is a chromosomal break at its locus and a subsequent fusion with another gene. We will follow the description of the two-groups procedure in Section 3.3.

## 4.2 Scores corresponding to principal components

Assume the $110 \times 9577$ matrix $\mathbf{X} = \begin{pmatrix} \mathbf{X}^{(1)} \\ \mathbf{X}^{(2)} \end{pmatrix}$. The weight vectors $\mathbf{d}_1, \mathbf{d}_2, \mathbf{d}_3, \ldots$ of the principal components are defined as the eigenvectors of the eigenvalue problem

$$(\mathbf{X} - \overline{\mathbf{X}})'(\mathbf{X} - \overline{\mathbf{X}})\mathbf{d}_h = \mathbf{d}_h \lambda_h, \quad \mathbf{d}'_h \mathbf{d}_h = 1, \quad h = 1, 2, 3, \ldots.$$

The eigenvalues $\lambda_1, \lambda_2, \lambda_3, \ldots$ are supposed to be ordered decreasingly. For efficient calculation, we use the "dual" eigenvalue problem with the smaller dimension $n = 110$

$$(\mathbf{X} - \overline{\mathbf{X}})(\mathbf{X} - \overline{\mathbf{X}})'\mathbf{z}_h = \mathbf{z}_h \lambda_h, \quad \mathbf{z}'_h \mathbf{z}_h = \lambda_h, \quad h = 1, 2, 3, \ldots.$$

The eigenvectors of the two eigenvalue problems are linked by the equations

$$\mathbf{z}_h = \pm(\mathbf{X} - \overline{\mathbf{X}})\mathbf{d}_h, \quad h = 1, 2, 3, \ldots.$$

The uncertainty of the signs in these equations is without importance, because we are only interested in two-sided score tests. Furthermore, we see that the eigenvectors $\mathbf{z}_h$ of the dual eigenvalue problem directly provide the score vectors that we need in our tests for sphericity. When solving the eigenvalue problems, the group labels of the patients, with or without myc-break, are not used.

The concrete sample of gene expression data yields the eigenvalues

$$\lambda_1 = 31198, \quad \lambda_2 = 24566, \quad \lambda_3 = 9676, \quad \lambda_4 = 7724, \ldots.$$

The corresponding beta tests for comparing the patients with and without myc-break, which are also the tests for sphericity, result in

$$B_1 = 0.1521, \quad B_2 = 0.3572, \quad B_3 = 0.0603, \quad B_4 = 0.0072, \ldots$$

with the *P*-values

$$P_1 = 2.5\text{E}-5, \quad P_2 = 5.4\text{E}-12, \quad P_3 = 0.0097, \quad P_4 = 0.3793, \ldots.$$

For $\alpha = 0.05$, these testing results prove that the principal component scores $z_1, z_2, z_3$ are non-spherically distributed and, accordingly, that differences between the groups (1) and (2) of patients are present. However, the testing procedure stops at the forth score, because of non-significance.

The tests for spherical distribution of the principal component scores are entirely correct in the mathematical sense, but they provide hardly any knowledge on the influence of single variables or sets of variables. Therefore, a second, more informative application of our new testing strategy is treated in the following.

### 4.3 Scores corresponding to data-based sets of variables

In this section, we will use the method in a more complicated context. Specifically, we combine multivariate tests with selection of variables. We consider as above the $110 \times 9577$ matrix $\mathbf{X} = (\mathbf{x}_1 \quad \mathbf{x}_2 \quad \ldots \quad \mathbf{x}_{9577})$ of the gene expression data. At first, a sequence of sets $m_1, m_2, m_3, \ldots$ is generated by collecting similar genes. Then for each gene set $m_h$, $h = 1, 2, 3, \ldots$, a corresponding score vector $\mathbf{z}_h$ is determined, and it is tested for sphericity of its distribution in the $(n-1)$-dimensional subspace. We emphasize that the gene sets $m_h$ and the score vectors $\mathbf{z}_h$ are constructed in a data-dependent way and that nevertheless exact tests for sphericity of the ordered score vectors $\mathbf{z}_1, \mathbf{z}_2, \mathbf{z}_3, \ldots$ are possible. The procedure has the property that all variations of the data $\mathbf{X}$ are taken into account, in particular the variations coming from randomly changing sets $m_h$.

We will again apply the method of generating gene sets that we have already published (Läuter, Horn, Rosolowski and Glimm, 2009). Each gene is considered as the centre of a gene set. The construction of the sets, the sorting, the selection and the calculation of scores are based on the sums of products matrix $(\mathbf{X} - \overline{\mathbf{X}})'(\mathbf{X} - \overline{\mathbf{X}})$. Therefore, this application runs also completely in the framework of our general concept, where score vectors $\mathbf{z}_h$ are tested for a spherical distribution.

A gene $i$ is included in the gene set $m$ of the centre gene $i_1$ if it has a smaller or an equal diagonal element in the matrix $(\mathbf{X} - \overline{\mathbf{X}})'(\mathbf{X} - \overline{\mathbf{X}})$, i.e.
$$(\mathbf{x}_i - \overline{\mathbf{x}}_i)'(\mathbf{x}_i - \overline{\mathbf{x}}_i) \leq (\mathbf{x}_{i_1} - \overline{\mathbf{x}}_{i_1})'(\mathbf{x}_{i_1} - \overline{\mathbf{x}}_{i_1}),$$
and if its correlation coefficient is not smaller than $\sqrt{0.5} \approx 0.7071$, i.e.
$$r_{i_1 i} = \frac{(\mathbf{x}_{i_1} - \overline{\mathbf{x}}_{i_1})'(\mathbf{x}_i - \overline{\mathbf{x}}_i)}{\sqrt{(\mathbf{x}_{i_1} - \overline{\mathbf{x}}_{i_1})'(\mathbf{x}_{i_1} - \overline{\mathbf{x}}_{i_1}) \cdot (\mathbf{x}_i - \overline{\mathbf{x}}_i)'(\mathbf{x}_i - \overline{\mathbf{x}}_i)}} \geq \sqrt{0.5}.$$
For each gene set arising in this way, the covariance measure
$$O_m = (\mathbf{x}_{i_1} - \overline{\mathbf{x}}_{i_1})'(\mathbf{x}_{i_1} - \overline{\mathbf{x}}_{i_1}) \sum_{i \in m} r_{i_1 i}$$
is determined, with the supplementary modification that, in the case of more than 20 partner genes $i$, the sum of the correlations is restricted to the 20 largest values $r_{i_1 i}$. Then, the gene sets $m$ are ordered by decreasing measures $O_m$. Beginning with the set of the highest value $O_m$ and continuing with the next following sets, a gene set is deleted from the sequence if it has a joint gene with any preceding, retained gene set. As an additional condition, we restrict the examination for joint genes to the 20 dominantly correlated genes in each set. The deletion step as a part of the gene-set generation process may be performed in our new concept of sphericity testing. This is an important improvement to our former method in Läuter et al. (2009), where mean values were tested.

Thus, an ordered sequence of gene sets $m_1, m_2, m_3, \ldots$ is obtained from the $p$ genes of the data matrix $\mathbf{X}$. The gene sets in the sequence decrease in their sizes, variances and correlations. Moreover, the gene sets contain no or only few joint genes. For each gene set $m_h$, $h = 1, 2, 3, \ldots$, a corresponding weight vector $\mathbf{d}_h$ is defined by

$$d_{ih} = \frac{1}{\sqrt{(\mathbf{x}_i - \overline{\mathbf{x}}_i)'(\mathbf{x}_i - \overline{\mathbf{x}}_i)}} \quad \text{for } i \in m_h, \qquad d_{ih} = 0 \quad \text{for } i \notin m_h.$$

Then, the standardized score vectors $\mathbf{z}_h = (\mathbf{X} - \overline{\mathbf{X}})\mathbf{d}_h$ can be calculated.

This method provides a data-driven, easily interpretable factorization of the multivariate information. The difference to the "anonymous" principal components considered before is evident.

For these gene sets and the corresponding score vectors, the beta tests for sphericity of the distributions are performed by comparing the patients with and without myc-break. The multiple procedure with the usual beta tests on the level $\alpha = 0.05$ yields significance only for score $\mathbf{z}_1$. The score $\mathbf{z}_2$ proves to be non-significant, so that the simple testing procedure ends here. However, the modified procedure by Hommel and Kropf with adjusted tests on the level $\alpha/40 = 0.00125$ and with stop after 40 non-significant results, provides 16 significant scores.

The following table shows the 16 significant score vectors $\mathbf{z}_1, \mathbf{z}_3, \mathbf{z}_4, \mathbf{z}_6, \ldots, \mathbf{z}_{52}$ obtained with the procedure by Hommel and Kropf. For each score vector, the Affymetrix ID of the centre gene of the set, the corresponding gene symbol, the gene name, the size of the set and the beta value of the two-groups test are presented. The shown gene sets overlap with signatures found by Hummel et al. (2006).

```
Score      Centre gene of the set                                    Size       Beta
num.
  1      205681_at      BCL2A1   BCL2-related protein A1             46 genes   0.6075
  3      203915_at      CXCL9    chemokine (C-X-C motif) ligand 9    17 genes   0.1534
  4      203680_at      PRKAR2B  protein kinase, cAMP-dependent      11 genes   0.3796
  6      204971_at      CSTA     cystatin A (stefin A)               21 genes   0.2700
  8      219148_at      PBK      PDZ binding kinase                 195 genes   0.1110
  9      212671_s_at    HLA-DQA1 major histocompatibility complex    13 genes   0.1858
 10      204249_s_at    LMO2     LIM domain only 2 (rhombotin-like 1) 6 genes   0.2879
 13      220532_s_at    LR8      NA                                  68 genes   0.2288
 23      209187_at      DR1      down-regulator of transcription 1   88 genes   0.0938
 30      218723_s_at    RGC32    NA                                   4 genes   0.4055
 31      208754_s_at    NAP1L1   nucleosome assembly protein 1-like 1 31 genes  0.1550
 33      206082_at      HCP5     HLA complex P5                      10 genes   0.2662
 39      209995_s_at    TCL1A    T-cell leukemia/lymphoma 1A          1 gene    0.1213
 40      201516_at      SRM      spermidine synthase                 17 genes   0.1268
 45      210072_at      CCL19    chemokine (C-C motif) ligand 19      3 genes   0.1850
 52      212592_at      ENAM     enamelin                             1 gene    0.2260
```

These are the results of our mathematically controlled procedure for model choice and selection of essential gene sets. Falsely significant scores – with a spherical distribution instead of a non-spherical one and not separating the groups of patients according to the myc-break – may appear with probability $\alpha = 0.05$, at most among the 16 gene sets based scores.

However, we should also note for the interpretation in the strict mathematical sense that such a procedure can provide sets of variables without mean-value differences between the groups (1) and (2) with a probability larger than $\alpha$. The derived scores are nevertheless non-spherically distributed in this case, and they discriminate the groups (1) and (2). We refer to the simulation results in Example 2 of Section 3.2, which demonstrate this seemingly contradictory behaviour.

Our former testing and selecting procedure (Läuter et al. (2009), Sections 5 and 6), which was based on the mean-value tests of Section 2.2, attains only 9 relevant, non-intersecting gene sets in this biostatistical example.

All presented methods are applicable in cases with a very large dimension *p*. They are characterized by simultaneously statistically securing the selection of variables and the deviation from the null hypothesis.

# References

Anderson, T.W. (1963). A test for equality of means when covariance matrices are unequal. *Annals of Mathematical Statistics* **34**, 671-672.

Anderson, T.W. (2003). *An introduction to multivariate statistical analysis. Third edition.* John Wiley & Sons.

Bai, Z. and Saranadasa, H. (1996). Effect of high dimension: an example of a two sample problem. *Statistica Sinica* **6**, 311-329.

Bennet, B.M. (1951). Note on a solution of the generalized Behrens-Fisher problem. *Annals of the Institute of Statistical Mathematics* **2**, 87-90.

Bentink, S. et al. (2008). Pathway activation patterns in diffuse large B-cell lymphomas. *Leukemia* **22** (9), 1746-1754.

Dempster, A.P. (1958). A high-dimensional two sample significance test. *Annals of Mathematical Statistics* **29**, 995-1010.

Dempster, A.P. (1960). A significance test for the separation of two highly multivariate small samples. *Biometrics* **16**, 41-50.

Fang, K.-T. and Zhang, Y.-T. (1990). *Generalized multivariate analysis.* Science Press, Beijing; Springer-Verlag Berlin, Heidelberg, New York, London, Paris, Tokyo, Hong Kong.

Goeman, J.J., van de Geer, S.A., de Kort, F. and van Houwelingen, H.C. (2004). A global test for groups of genes: testing association with a clinical outcome. *Bioinformatics* **20**, 93-99.

Hastie, T., Tibshirani, R. and Friedman, J. (2001). *The elements of statistical learning.* Springer, New York.

Hommel, G. and Kropf, S. (2005). Tests for differentiation in gene expression using a data-driven order of weights for hypotheses. *Biometrical Journal* **47**, 554-562.

Hotelling, H. (1931). The generalization of Student's ratio. *Annals of Mathematical Statistics* **2**, 360-378.

Hummel, M. et al. (2006). A biologic definition of Burkitt's lymphoma from transcriptional and genomic profiling. *The New England Journal of Medicine* **354**, 2419-2430.

Kropf, S. (2000). *Hochdimensionale multivariate Verfahren in der medizinischen Statistik.* Shaker, Aachen.

Kropf, S. and Läuter, J. (2002). Multiple tests for different sets of variables using a data-driven ordering of hypotheses, with an application to gene expression data. *Biometrical Journal* **44**, 789-800.

Langsrud, Ø. (2005). Rotation tests. *Statistics and Computing* **15**, 53-60.

Läuter, J. (1996). Exact t and F tests for analyzing studies with multiple endpoints. *Biometrics* **52**, 964-970.

Läuter, J., Glimm, E. and Kropf, S. (1996). New multivariate tests for data with an inherent structure. *Biometrical Journal* **38**, 5-23, Erratum: *Biometrical Journal* **40**, 1015.

Läuter, J., Glimm, E. and Kropf, S. (1998). Multivariate tests based on left-spherically distributed linear scores. *The Annals of Statistics* **26**, 1972-1988, Correction: *The Annals of Statistics* **27**, 1441.

Läuter, J., Horn, F., Rosolowski, M. and Glimm. E. (2009). High-dimensional data analysis: Selection of variables, data compression and graphics – Application to gene expression. *Biometrical Journal* **51**, 235-251.

Lehmann, E.L. (1975): *Nonparametrics: Statistical methods based on ranks,* John Wiley and Sons, New York,

Liang, J. and Fang, K.-T. (2000). Some applications of Läuter's technique in tests for spherical symmetry. *Biometrical Journal* **42**, 923-936.

Maneck, M., Schrader, A., Kube, D. and Spang, R. (2011). Genomic data integration using guided clustering. *Bioinformatics* **27** (16), 2231-2238.

Meinshausen, N. (2008). Hierarchical testing of variable importance. *Biometrika* **95**, 265-278.

Rao, C.R. (1948). Tests of significance in multivariate analysis. *Biometrika* **35**, 58-79.

Rao, C.R. (1973). *Linear statistical inference and its applications,* 2nd ed., John Wiley & Sons, New York.

Roy, S.N. (1953). On a heuristic method of test construction and its use in multivariate analysis. *Annals of Mathematical Statistics* **24**, 220-238.

Schramedei, K., Mörbt, N., Pfeifer, G., Läuter, J., Rosolowski, M., Tomm, J.M., von Bergen, M., Horn, F. and Brocke-Heidrich, K. (2011). MicroRNA-21 targets tumor suppressor genes ANP32A and SMARCA4. *Oncogene* **30**, 2975-2985.

Seber, G.A.F. (2004). *Multivariate observations.* John Wiley & Sons, Hoboken, New Jersey.

Srivastava, M.S. and Du, M. (2008). A test for the mean vector with fewer observations than the dimension. *Journal of Multivariate Analysis* **99**, 386-402.

Timm, N.H. (2002). *Applied multivariate analysis.* Springer-Verlag, New York.

Westfall, P.H. and Young, S.S. (1993). *Resampling-based multiple testing.* John Wiley & Sons, New York.

Wilks, S.S. (1932). Certain generalizations in the analysis of variance. *Biometrika* **24**, 471-494.

Yao, Y. (1965). An approximate degrees of freedom solution to the multivariate Behrens-Fisher problem. *Biometrika* **52**, 139-147.